\documentstyle[aas2pp4,epsf]{article}
\received{}
\accepted{}
\begin{document}
\def\etal{{et al. \rm}}
\def\l{$\lambda$}
\def\p{$\phi$ $\sim$ }
\def\km{km s$^{-1}$}
\title{A 2.3 DAY PERIODIC VARIABILITY IN THE APPARENTLY
SINGLE WOLF-RAYET STAR WR 134: COLLAPSED COMPANION OR ROTATIONAL MODULATION?}
\author{Thierry Morel,\altaffilmark{1,}\altaffilmark{*} Sergey
V. Marchenko,\altaffilmark{1} Philippe R. J. Eenens,\altaffilmark{2}
Anthony F. J. Moffat,\altaffilmark{1,}\altaffilmark{3} Gloria
Koenigsberger,\altaffilmark{4} Igor
I. Antokhin,\altaffilmark{5} Thomas
Eversberg,\altaffilmark{1,}\altaffilmark{**} Gaghik H.
Tovmassian,\altaffilmark{6} Grant M. Hill,\altaffilmark{7} Octavio
Cardona,\altaffilmark{8} and Nicole St-Louis\altaffilmark{1}}
\altaffiltext{1}{D\'epartement de Physique, Universit\'e de Montr\'eal,
C.P. 6128, Succ. Centre-Ville, Montr\'eal, Qu\'ebec, Canada, H3C 3J7;
and Observatoire du Mont M\'egantic; morel, sergey, moffat, eversber, stlouis@astro.umontreal.ca.\\
$\star$ Present adress: Astrophysics Group, Imperial College of Science, Technology and Medicine, Blackett Laboratory, Prince Consort Road, London, SW7 2BZ, UK; email: morel@ic.ac.uk.\\
$\star$$\star$ Present adress: Feinfocus Medizintechnik GmbH, Im Bahlbrink 11-13, 30827, Garbsen, Germany; email: t\_eversberg@feinfocus.com.}
\altaffiltext{2}{Departamento de Astronom\'{\i}a, Universidad de Guanajuato, 
 Apdo. Postal 144, 36000 Guanajuato Gto, M\'exico; eenens@carina.astro.ugto.mx.}
\altaffiltext{3}{Killam Research Fellow of the Canada Council for the Arts.}
\altaffiltext{4}{Instituto de Astronom\'{\i}a, UNAM, Apdo. Postal 70-264,
 M\'exico D.F. 04510, M\'exico; gloria@astroscu.unam.mx.}
\altaffiltext{5}{Sternberg Astronomical Institute, Universiteskij
Prospect 13, 119899, Moscow, Russia; igor@sai.msu.su.}
\altaffiltext{6}{Instituto de Astronom\'{\i}a, Apdo. Postal 877,
C. P. 22860, Ensenada, B. C., M\'exico; gag@bufadora.astrosen.unam.mx.}
\altaffiltext{7}{McDonald
Observatory, HET, P. O. Box 1337, Fort Davis, TX; grant@astro.as.utexas.edu.}
\altaffiltext{8}{Instituto Nacional de Astrof\'{\i}sica, Optica y
Electr\'onica, Apdo. Postal 51, Puebla, Pue. 72000, M\'exico; ocardona@inaoep.mx.}
\slugcomment{Article submitted to the Astrophysical Journal main
section.}
\begin{abstract}
The apparently single WN 6 type star WR 134 (HD 191765) is distinguished
among the Wolf-Rayet star population by its strong, presumably
cyclical ($\cal P$ $\approx$ 2.3 day), spectral variations. A true
periodicity --- which is still very
much debated --- would render WR 134 a prime candidate for harboring
either  a collapsed companion or a rotating, large-scale, inhomogeneous outflow.

We have carried out an intensive campaign of spectroscopic
and photometric monitoring of WR 134 from 1989 to 1997
in an attempt to reveal the true nature of this object. This
unprecedentedly large data set allows us to confirm unambiguously the
existence of a coherent 2.25 $\pm$ 0.05 day
periodicity in the line-profile changes of He II \l4686, although the
global 
pattern of variability is different from one epoch to another. This period is only marginally detected in the photometric
data set.

Assuming the 2.25 day periodic variability to be induced by orbital
motion of a collapsed companion, we develop a simple model aiming to
investigate (i) the effect of this strongly
ionizing, accreting companion on the Wolf-Rayet wind structure, and (ii) the
expected emergent X-ray luminosity. We argue that the predicted and
observed X-ray fluxes can only be matched if the accretion on the
collapsed star is significantly
inhibited. Additionally, we performed simulations of line-profile variations
caused by the orbital revolution of a localized, strongly ionized wind cavity surrounding the X-ray source. A reasonable fit is
achieved between the observed and modeled
phase-dependent line profiles of He II \l4686. However, the
derived size of the photoionized zone substantially exceeds our
expectations, given the observed low-level X-ray flux.

Alternatively, we explore rotational
modulation of a persistent, largely
anisotropic outflow as the origin of the observed cyclical
variability. Although qualitative, this hypothesis leads to greater
consistency with the observations.
\end{abstract}
\it Subject headings: \rm stars: individual (WR 134) --- stars: mass
loss --- stars: Wolf-Rayet 
\section{Introduction}
The recognition in the 1980s that some apparently single Wolf-Rayet (WR)
stars exhibit seemingly periodic line profile and/or photometric variations
argued for the existence
of systems made up of a WR star and a collapsed companion (hereafter WR + $c$; where $c$
stands either for a neutron star or a black hole), as predicted by the general theory of massive close-binary evolution (e.g., van den Heuvel
\& de Loore 1973): O + O $\rightarrow$ WR + O $\rightarrow$ $c$ + O
$\rightarrow$ $c$ + WR $\rightarrow$ $c$ + $c$. Due to the recoil of the
first supernova explosion leading to $c$ + O, the system generally acquires a high systemic
velocity. If massive enough, the
secondary evolves in turn into a WR star, at which point the system may
have reached an unusually high galactic
latitude for a population I star. These two peculiarities (runaway
velocity and position), along with the existence of a surrounding ring nebula (presumably
formed by matter ejected during the secondary mass exchange), were among the criteria initially used to
select WR + $c$ candidates (van den Heuvel 1976; Moffat 1982;
Cherepashchuk \& Aslanov 1984).\footnote{Since then, only Cygnus X-3 has
been shown to probably belong to this class (van Kerkwijk \etal
1996; Schmutz, Geballe, \& Schild 1996; but see Mitra 1998). Besides this system, the
single-line WN 7 star WR 148 (HD 197406) constitutes one of the most promising
candidates (Marchenko \etal 1996a). Two
candidates for WR + $c$ in 30 Doradus have also been proposed by Wang (1995).} 

A major breakthrough in the qualitative scenario described above comes
from a recent redetermination of the
distribution of the radio pulsar runaway velocities. In sharp contrast with earlier estimations (100-200 \km; Gunn \& Ostriker 1970; Lyne,
Anderson, \& Salter 1982), these new values imply a mean pulsar kick
velocity at birth of about 450 \km \ (Lyne \& Lorimer 1994; Lorimer,
Bailes, \& Harrison 1997). Although such a high kick velocity imparted
at birth is still much debated (e.g., Hansen \& Phinney 1997; Hartman
1997), this result tends to show --- as suggested  by the apparent paucity of ``runaway'' OB stars with
compact companions (Kumar, Kallman, \& Thomas 1983; Gies \& Bolton 1986; Philp \etal 1996;
Sayer, Nice, \& Kaspi 1996) --- that the number of systems which would survive the
first supernova explosion is probably considerably lower than initially thought (see Brandt \&
Podsiadlowski 1995 vs Hellings \& de Loore 1986). Indeed, the incorporation of up-to-date physics of supernova
explosions in the most recent population synthesis models of massive binaries
leads to a very small number of observable WR + $c$ systems in the Galaxy
($N$ $\la$ 5; De Donder, Vanbeveren, \&
van Bever 1997). This number contrasts significantly with the number of
 observationally selected candidates ($N$ $\approx$ 15; Vanbeveren 1991),
especially considering the necessarily incomplete nature of this sample. At face value, this
discrepancy might imply that most of these WR + $c$ candidates are spurious or, more
interestingly, that other physical mechanisms are at work to
induce the
large-scale, periodic variability inherent to \it some \rm objects. 

At least in O stars, the progenitors of WR stars, it appears that large-scale periodic
variability may be induced by rotating aspherical and structured winds
(e.g., Fullerton \etal 1997; Kaper \etal 1997). The incidence of asymmetric outflows among the
WR star population is probably lower, although recent spectropolarimetric
(Harries, Hillier, \& Howarth 1998), UV (St-Louis \etal 1995), optical (Marchenko \etal 1998a), and radio observations (White \&
Becker 1996; Williams
\etal 1997) have now independently revealed this peculiarity
in a substantial number of WR stars. In order to account for
the periodic variability inferred for the suspected WR + $c$
candidates, the rotational modulation of a persistent, largely
inhomogeneous outflow could thus constitute in some cases an attractive alternative to the binary
scenario, and would be consistent with the lack of strong, accretion-type X-ray emission
(Wessolowski 1996).

The nature of the variable WN 5 star EZ CMa (HD 50896; WR 6), which is generally
considered as a prime candidate for an orbiting collapsed
companion (Firmani \etal 1980), has been recently
re-investigated by means of extensive spectropolarimetric (Harries \etal 1999), UV (St-Louis
\etal 1995), and
optical studies (Morel, St-Louis, \& Marchenko
1997; Morel \etal 1998). These studies reveal that  the 3.77 day period displayed by
this object is more likely induced by the rotational modulation of a
structured wind. Hints of long-term (from weeks to months) wind variability triggered by a spatial dependence of the physical
conditions prevailing in the vicinity of the hydrostatic stellar ``surface'' were also
found (Morel \etal 1998).

Besides the search for WR + $c$ systems, the mysterious origin of such deep-seated
variability (possibly induced by pulsational instabilities or large-scale magnetic
structures in the case of EZ CMa) has prompted us to carry out intensive
spectroscopic and photometric observations of other WR stars suspected to
exhibit such periodic line-profile
variations (LPVs). In this context, WR 134 is regarded as a natural target.
\section{Observational Background}
WR 134 is a relatively bright ($v$ $\approx$ 8.3) WN 6 star
surrounded by a ring-type nebula embedded
in the H II region S 109 (Crampton 1971; Esteban \& Rosado 1995). 

Following the discovery of spectacular line-profile and photometric
changes intrinsic to this object (Bappu 1951; Ross 1961), a major
effort was directed to establish the periodic nature of these variations.
Cherepashchuk (1975) was the first to investigate photometric
variability of WR 134 on
different timescales. Irregular night-to-night light
fluctuations (with rms amplitude $\sigma \approx$ 0.014 mag) were
observed, without evidence
for rapid ($\sim$ hourly) or long-term ($\sim$ monthly)
changes. The first claim
of periodic variability with $\cal P$ = 7.44 $\pm$ 0.10 day was made by
Antokhin, Aslanov, \& Cherepashchuk (1982) on the basis of a two-month
interval of broadband
photometric monitoring. The authors
also reported small amplitude
radial velocity variations present in simultaneously acquired
spectroscopic data ($K_{WR}$ $\approx$ 20-40 \km), consistent
with the above period. This period was subsequently improved to $\cal P$ =
7.483 $\pm$ 0.004 day
by Antokhin \& Cherepashchuk (1984). Zhilyaev \& Khalack (1996) established
that the short-term stochastic variability presented by WR 134 is not related to this
period, as it would be in the case of an orbiting collapsed companion. Moffat \&
Shara (1986) were unable to identify this period in their photometric
data, although they found evidence
for a 1.8 day periodicity, first tentatively reported by Lamontagne (1983)
on the basis of radial velocity measurements. Following the recognition
that the level of continuum flux from WR 134 is also irregularly variable on a timescale of weeks
to months (Antokhin \& Volkov 1987), periodic changes of the
equivalent widths of He II \l4859 with $\cal P$ = 1.74 $\pm$ 0.38 day were reported by
Marchenko (1988). Then, Robert (1992) reported the
existence of a 2.34 day periodicity in radial velocity variations, which
could be a one-day
alias of the 1.8 day period claimed in previous studies. A two-day
quasi-periodicity in photometric
data was also
reported by Antokhin \etal (1992). Although not obvious in
the spectroscopic data set obtained by Vreux \etal (1992) because of an unfortunate correspondence to a badly sampled frequency domain (see also
Gosset, Vreux, \& Andrillat 1994; Gosset
\& Vreux 1996), strong hints toward the
reality of this two-day recurrence timescale come from the analysis of an extensive set of
optical spectra by McCandliss \etal (1994), who derived a 2.27 $\pm$ 0.04 day
periodicity in the centroid, second moment and skewness
of the line-profile time series. A relatively large-scale, likely cyclical, shift of
excess-emission components superposed on the underlying line
profiles was also observed.

WR 134 also shows strong variations in broadband  polarimetry (Robert \etal
1989) and spectropolarimetry (Schulte-Ladbeck \etal 1992).

Numerous models have been put forward to account for the intricate
variability pattern, including: the presence of an orbiting
neutron star (Antokhin \& Cherepashchuk 1984), a disk connected to
the central star by ever-changing filaments (Underhill
\etal 1990), or a bipolar magnetic outflow (Vreux \etal 1992).

No consensus has been reached yet concerning the possible periodic
nature of the variations in WR 134. We present in this paper the analysis of
a long-term campaign of optical photometric and spectroscopic monitoring, in an
attempt to shed new light on this issue. Preliminary results concerning the photometric
data subset were presented by Moffat \& Marchenko (1993a, b) and Marchenko
\etal (1996b).
\section{Observations and Reduction Procedure}
\subsection{Spectroscopy}
WR 134 was observed spectroscopically between 1992 and 1995. A
journal of observations is presented in Table 1, which lists an epoch number, the dates of the spectroscopic observations, the interval of the observations
in Heliocentric Julian Days, the
 observatory name, the number of CCD spectra obtained, the selected spectral domain, the reciprocal dispersion of the spectra, and the typical
signal-to-noise ratio (S/N) in the continuum. The spectra were reduced using the {\tt IRAF}\footnote{{\tt IRAF} is
distributed by the National Optical Astronomy Observatories, operated by
the Association of Universities for Research in Astronomy, Inc., under
cooperative agreement with the National Science Foundation.} data reduction
packages. Bias subtraction, flat-field division, sky subtraction,
extraction of the spectra, and wavelength calibration were carried out in
the usual way. Spectra of calibration lamps were taken immediately before, and after the stellar exposure. The rectification of the
spectra was carried out by an appropriate choice of line-free
regions, subsequently fitted by a low order Legendre
polynomial. 

In order to minimize the spurious velocity shifts induced by
an inevitably imperfect wavelength calibration, the spectra were
coaligned in velocity space by using the interstellar doublet Na I \l\l5890, 5896, or the diffuse interstellar band at 4501
\AA \ 
as fiducial marks.

Each spectrum obtained during epochs I and II generally consists of the sum of
three consecutive exposures typically separated by about 10 minutes, with 
no apparent short-term variability.  This is expected in view of
the longer timescales required to detect any
significant motion of \it stochastic, small-scale \rm emission-excess features travelling on
top of the line profiles induced by outwardly moving wind
inhomogeneities (L\'epine \& Moffat 1999). 

No attempts
have been made to correct for the continuum level variability, owing
to its relatively low amplitude and irregularity ($\S$ 4.1.2).
\subsection{Photometry}
Previous work led us to expect a fairly complex light curve
behavior. Thus, in attempting to reveal
regular (i.e., periodic) variations, it was deemed essential to combine
multi-epoch observations of high quality. We therefore organized a
long-term campaign of $UBV$ photometric
monitoring  of WR 134 in 1990-1997 using the 0.25 m Automatic
Photometric Telescope (APT; Young \etal 1991). The APT data collected during the
first 3 years of monitoring have  been discussed by Moffat \& Marchenko
(1993a, b). The data generally consist of systematic, $\sim$ day-to-day
photometry with 1-2 measurements per night, each of accuracy of $\sigma$ $\approx$ 0.005-0.008 mag. HD
192533 and HD 192934 were used as check and comparison stars,
respectively.

 In 1992 and 1995, we also organized additional multi-site, very
intensive photometric campaigns, in an attempt to support the simultaneous spectroscopy. In 1992, we used the 0.84 m telescope of San Pedro M\'artir
Observatory (Mexico) and the 0.6 m telescope of Maidanak
Observatory (Uzbekistan, fSU), to observe WR 134 in narrow and broadband
filters. In Mexico, the star was observed with a two-channel (WR
and guide star) photometer in  rapid (5 s resolution) or
ultra-rapid (0.01 s; see Marchenko \etal 1994) mode, through filters centered at $\lambda _c$
= 5185 \AA \ (FWHM = 250 \AA, $\sim$ stellar continuum), and 
$\lambda _c$ = 4700 \AA \ (FWHM = 190 \AA, He II $\lambda$4686 emission
line). By subsequently re-binning the rapid-photometry data into 0.1 day
bins, we achieved an accuracy of 0.002-0.003 mag per bin. Additionally, we used the two-channel device as a conventional
one-channel photometer to obtain differential $UBV$ observations 1-2
times per night (individual accuracy of
0.006-0.007 mag), with  HD 192533 and HD 192934 as a check-comparison
pair. The single-channel photometer of Maidanak Observatory was equipped with
 a single $\lambda _c$ = 6012 \AA \ (FWHM = 87 \AA, $\sim$ stellar
continuum) filter. HD 191917 was used as a comparison star, resulting in
a  mean accuracy of 0.003 mag per 0.1 day binned data point. 

In 1995, we used the 0.84 m telescope of San Pedro M\'artir
Observatory and one-channel photometer with a broadband $V$ filter in
differential photometry mode, providing a typical accuracy of
0.003-0.005 mag per 0.1 day binned point. WR 134 was also monitored by the Crimean Observational Station of GAISH (Crimea, Ukraine) with a
one-channel photometer ($V$ filter) attached to a 0.6 m telescope, providing a mean accuracy of 0.007 mag per 0.1 day bin. 

We also used all the photometric data secured by the $HIPPARCOS$
astrometric satellite in 1989-1993 (broadband $Hp$ system; see Marchenko \etal 1998b). To match the accuracy of the ground-based observations, 
we rebinned the $HIPPARCOS$ data set to 1 day bins, achieving $\sigma$
$\approx$ 0.006 mag per
combined data point. All relevant information concerning the photometry
is provided in
Table 2.
\section{Results}
\subsection{Search for Short-Term Periodicity}
\subsubsection{Spectroscopy}
Inspired by the previous clear indication of a 2.3 day periodicity in the
 line-profile variations found by McCandliss \etal (1994), we have
performed a similar search in the present data set by calculating the power spectra
(PS) using the technique of Scargle (1982) on the skewness,
centroid, and FWHM time series of He
II \l4686, the strongest line. The centroid and skewness were calculated
as the first moment, and the ratio of the third and the (3/2
power of the) second central
moments of the line profile, respectively. In order to minimize the
contributions of blends, both measurements were restricted to the portion of the profile above two in units of the continuum. The FWHM was determined by a Gaussian
fit to the entire line profile. A subsequent correction of the frequency
spectrum by the CLEAN
algorithm was performed in order to remove aliases and spurious features
induced by the unevenly spaced nature of the data (see Roberts, Leh\'ar, \&
Dreher 1987). The period search was
performed up to the Nyquist frequency for unevenly spaced
data: $\nu_N = (2\Delta t_{min})^{-1}$. However, since no evidence for periodic signals was found at high frequencies, we will only display in the following the PS for frequencies up to $\nu_N = (2\Delta t_{mean})^{-1}$. The data
acquired during epoch III suffer from large time gaps and are
inadequate in the search for periods of the order of days. Also, because of the larger time span of epoch I compared to epoch II (and of the related higher frequency
resolution), the period determination
was mainly based on the former data set. The ``raw'' and CLEANed PS of
the skewness, centroid, and FWHM time series of He II \l4686 are
presented in Figure 1. Note that the centroid variations are mainly due to
changes in the line profile morphology and thus do \it not \rm
reflect a global shift of the profile. The significance of the peaks in the PS can be estimated by
means of the 99.0 \% and 99.9 \%
thresholds (Fig.1), giving the
probability that a given peak is related to the presence of a deterministic signal in the time series
(Scargle 1982). Note that these levels are only indicative in the case of
unevenly spaced data (see, e.g., Antokhin \etal 1995). 

The ``raw'' PS display numerous
peaks, most of them being induced by the fairly complex
temporal window and/or the presence of noise. Very few significant peaks remain in the CLEANed PS. In
particular, the
highest peaks in the skewness and centroid PS are found at $\nu_0$ $\approx$
0.442 and 0.447 d$^{-1}$ (with a typical uncertainty of about 0.010 d$^{-1}$, as
determined by the FWHM of the peaks), which translates into periods of
2.26 $\pm$ 0.05 and 2.24 $\pm$ 0.05
day, respectively. These values are equal, within the uncertainties, to the period proposed by McCandliss \etal (1994): 2.27 $\pm$ 0.04 day. The PS for the FWHM
time series presents a prominent peak at $\nu_1$ $\approx$ 0.884
d$^{-1}$, which can very likely be identified with the first harmonic of the
``fundamental'' frequency $\nu_0$ suggested above (we adopt $\nu_0$ = 0.444
d$^{-1}$ in the following). A highly significant
signal is also found at
$\nu_2$ $\approx$ 0.326 d$^{-1}$ ($\cal P$ = 3.07 $\pm$ 0.10 day); we will return to this point below. Note the absence of any trace of the 7.483 day
period ($\nu$ $\approx$ 0.134 d$^{-1}$) proposed by Antokhin \& Cherepashchuk (1984).
The periodic nature of the skewness, centroid, and FWHM variations with a frequency of $\nu_0$ is
confirmed --- at least for epochs I and II --- when the data are plotted as a
function of phase (Fig.2).\footnote{Because
our period of 2.25 $\pm$ 0.05 day is indistinguishable from the one derived by McCandliss \etal (1994), we adopt their ephemeris in the following:
$HJD$ 2,447,015.753 + 2.27 $E$.}
We have also calculated the equivalent widths (EWs) of He II \l4686 by
integrating the line flux in the interval 4646-4765 \AA. Although
significant, the EW variations show no clear
evidence for phase-locked variability (Fig.2).

After demonstrating that the integrated
properties of the line profile are periodic in nature, it is natural to ask
whether the same recurrence timescale is present in the detailed LPVs. To this end, we performed
pixel-to-pixel CLEANing of the epoch I and II data sets. All
spectra were rebinned in a similar manner and were used as
independent time sequences, thus creating 1024 CLEANed PS along the wavelength direction. This procedure confirms the presence of $\nu_0$, along with its first harmonic at
$\nu_1$ $\approx$ 0.88 d$^{-1}$ over the entire line profile of He II
\l4686 during epoch I, along with weak traces of this period in He II \l4542, N V \l\l4604, 4620 and 
N III \l4640 (Fig.3). As in the PS of the
FWHM time series of epoch I (Fig.1), a periodic signal at $\nu_2$
$\approx$ 0.326 d$^{-1}$ is also present in the pixel-to-pixel CLEANed PS. The
occurence of the same signal at $\nu_2$ in the FWHM and pixel-to-pixel CLEANed PS raises the possiblity that this feature is real. However, $\nu_0$
(or its harmonics) is generally (within the
uncertainties) recovered in the epoch II
PS, contrary to $\nu_2$. More importantly, the gray-scale plot of
He II \l4686 does not show a coherent pattern of
variability when folded with $\nu_2$, whereas it does when folded with
$\nu_0$ (see below). Also, the phase diagram of the FWHM
data of epoch I is considerably noisier
when folded with this frequency. Further observations are essential to indicate whether
this period is genuinely spurious. Although the LPVs of He II \l4686
for epoch II are undoubtedly periodic with a frequency of $\nu_0$
(see below), only a weak signal appears in the pixel-to-pixel
PS at this frequency on the top of He II \l4542 and He II \l4686. This may be related to
the more complex pattern of variability compared to epoch I.

The next analysis we have performed consisted in grouping the spectra for each epoch into 0.02 phase
bins. The individual, binned He II \l4686 line profiles minus their
corresponding unweighted means for each epoch are arranged as a function
of phase in the upper panels of Figures 4a-4c. This line transition was chosen
because its LPVs are quite representative of (but stronger
and clearer than) that of the other He II
spectral features ($\S$ 4.3). In view of the long time span of the
observations (13, 4, and 63 cycles for epochs I, II, and III,
respectively), He II \l4686 presents a remarkably coherent
phase-related pattern of variability. However, it is
noteworthy that significant cycle-to-cycle differences in the  line-profile morphology are often found, a fact which is reflected in the gray-scale plots
by discontinuities in the pattern of variability between consecutive
bins (especially for epoch III; Fig.4c). Four main factors can induce this: (i) lack of long-term
coherency in the pattern of variability; (ii) artificial loss of coherency induced
by the uncertainty in the period; (iii) presence of additional
small-scale profile variations created by shocked (and possibly
turbulent) material carried out by the
global stellar outflow, as observed in other WR stars (L\'epine \& Moffat 1999); and (iv) continuum flux
variations for which no allowance has been made; we deem
this last factor as a less important effect because of the small amplitude of the
changes (see below). Concerning the first
two points, note that the smaller the time span of the
observations (Table 1), the
higher the coherency.
\subsubsection{Photometry}
For the analysis of the broadband photometry, we only use the $V$-filter data as being less contaminated by emission lines (about 7 \% of the 
total $V$ flux), along with all available mediumband visual-filter photometry, 
in total 734 observations binned to 0.1 day (1 day for the $HIPPARCOS$
data).  Note that for WR 134, the small-scale variations observed in $U$, $B$, and $V$  are fairly well-correlated in general (Moffat \& Marchenko 1993b).
  
Combining all 1989-1997 data and reducing them to the APT photometric
system, we concentrate on a search for relatively short periods; a
detailed investigation of the temporal behavior of the
``secular'' components (i.e., $\cal P$ $\approx$ 625 day and 
 $\cal P$ $\approx$ 40 day variations) being presented in  Marchenko \etal
(1996b) and Marchenko \& Moffat (1998a). 

We constructed a CLEANed PS of the whole photometric data set,
but failed to find any significant peaks in the range of interest, i.e.,
around the expected 2.3 day component. This is possibly due to the lack
of coherency of the period on long timescales. The long-term component with
$\cal P$ $\approx$ 625 day completely dominates  the PS, followed by a less
evident power peak at $\cal P$ $\approx$ 40 day. Pre-whitening from both
long-period variations does not improve the situation in the high-frequency domain. 

In the analysis of  the data subsets, we have encountered some limited
success. The ``subsets'' are naturally-imposed segments (e.g., summer monsoon and 
winter gaps at the APT, short-term campaigns of intense monitoring), with gaps between 
the segments exceeding, or comparable to, the length of the segment filled by 
observations. We obtain 14 such segments, covering the period HJD 2,447,859-2,450,636, with the number of observations per segment varying from 9 to 161.
The most frequently appearing detail in the CLEANed PS is  clustered 
around $\nu$ = 1.3-1.4 d$^{-1}$ (in 5 out of 14 PS). Twice, we detect 
a fairly strong signal at $\nu$ $\approx$ 0.9 d$^{-1}$: HJD 2,448,896-2,448,976 (43 observations), and
HJD 2,450,522-2,450,636 (82 observations). Concerning the most abundant data set secured 
during the 1992 campaign (161 data points distributed over 135 days; practically all of them are shown on 
Fig.5), the expected signal at $\nu_0$ = 0.440
$\pm$ 0.018 d$^{-1}$ is enhanced only during a few cycles, between 
HJD 2,448,818-2,448,824 (Fig.5). We have used the data shown in the middle section 
of Figure 5 to construct a folded light curve with $\cal P$
= 2.27 day (ephemeris from McCandliss \etal 1994). Despite the fact that the periodic signal found during the interval HJD 2,448,818-2,448,824 is identical to the one found in the
independently and practically simultaneously acquired spectroscopic
data set --- thus demonstrating that this detection is unlikely to be fortuitous ---, the very noisy appearance
of the folded light curve (Fig.6) shows that the detection of the $\cal P$
= 2.27 day periodicity is only marginal in the photometric data set. Adopting $\nu_0$ $\approx$ 0.44
d$^{-1}$ as the principal frequency, we may interpret the $\nu$ $\approx$ 0.9 and 1.3-1.4 d$^{-1}$ components found in the other subsets as the first and second harmonics, respectively. 

Another intensive multi-site monitoring campaign in 1995 reveals the presence of
the long-term $\cal P$ $\approx$ 40 day cyclic component (Marchenko et
al. 1996b, and our Fig.7), masking the relatively weaker $\nu_1$ $\approx$
0.88 d$^{-1}$ frequency,  along with the possible initiation of $\cal P$ $\approx$
7-8 day  variations (perhaps as in Antokhin \& Cherepashchuk 1984) at HJD
2,449,920-2,449,948. The latter phenomenon points to the transient character
of the $\cal P$ $\approx$ 7-8 day variations.

\subsection{Similarities of the Variations Across a Given Line Profile}
In an attempt to objectively and quantitatively examine a potential
relationship in the pattern of variability displayed by different portions
of a given line profile, we have calculated the Spearman rank-order correlation
matrices (see, e.g., Johns \& Basri 1995; Lago \& Gameiro 1998) whose elements $r(i,j)$ give the degree of correlation between the line
intensity variations at any pixels $i$ and $j$ across the line profile
(these matrices are symmetric and a perfect positive correlation is found along the main diagonal where $i$ =
$j$). These matrices for He II
\l4686 are shown for all epochs in Figure 8, in the form of contour
plots, where the
lowest contour indicates a significant positive or negative correlation at the 99.9 \%
confidence level.

An inspection of these matrices allows one to draw the following general
conclusions: (i) There are various regions in the auto-correlation
matrices presenting a significant positive or negative correlation; (ii) If the
variations were principally induced by the changes in the continuum flux
level ($\S$ 4.1.2), one would
observe a tendency for a positive correlation over a substantial fraction of velocity space; this is clearly not observed; (iii) The
main diagonal is not a straight line with a width corresponding to the
velocity resolution (at most 100 km s$^{-1}$ in our case), as expected if two contiguous velocity elements would vary in a completely uncorrelated fashion, but is
considerably wider. This suggests that the same pattern of
variability affects simultaneously
a fairly large velocity range of the profile, as independently found by L\'epine, Moffat, \& Henriksen
(1996), who derived a relatively large mean line-of-sight velocity
dispersion for the emission subpeaks
travelling across the line profile of WR 134 ($\overline{\sigma_{\xi}}$ $\approx$ 440 km
s$^{-1}$) compared to other WR stars in their sample ($\overline{\sigma_{\xi}}$ $\approx$ 100 km s$^{-1}$).

If we turn to a detailed epoch-to-epoch comparison of the matrices in
Figure 8, we find significant
differences (note that the phase coverage is sufficiently similar for
all epochs for a
direct comparison to be made). For example,
a positive correlation is observed for epoch I at
(-- 500, + 2400) and (+ 200, + 1800) \km, whereas the variations in these
velocity ranges are rather negatively correlated for epoch II. The paucity
of contours for epoch III indicates that no significant relationship
existed for this epoch between the
changes presented by different portions of the He II \l4686 line profile. This leads to
the conclusion that the global pattern of variability is likely to differ notably in nature on a yearly
timescale, although the LPVs are coherent over shorter timescales when
phased with the 2.27 day period (Figs.4a-4c). 
\subsection{Similarities Between the Variations of Different Lines}
The similar qualitative behavior of different lines has already been
emphasized in the past (McCandliss 1988; Marchenko 1988; Vreux \etal 1992; McCandliss \etal 1994). Here, we re-address this point by
calculating the degree of correlation between the LPVs at different projected velocities (referred to the line laboratory rest wavelength) in
two given line profiles. The spectra obtained during epoch III are well suited for this purpose because of
their wide spectral coverage. The correlation matrices of He II \l5412 with He II
\l4542, He II \l4686, He II \l4859, and the doublet C IV \l5806 are presented for this epoch in
Figure 9. As revealed by the
prevalence of
contours along the main diagonal, the LPVs presented by the He II lines
are generally well correlated (this is especially true for the
relatively unblended lines He II \l5412 and
He II \l4686), emphasizing that they vary in a fairly similar
fashion. This is remarkable as the presence of blends and/or
noise tends to mask any potential correlation. Only the blue wings of He
II \l5412 and C IV \l5806 are positively correlated for epoch III. The changes affecting He II \l4686 and He II \l4542 are also
well correlated  for epochs I and II.
\section{Discussion}
\subsection{The Duplicity of WR 134 Questioned}
In view of the long-suspected association of WR 134 with an orbiting
collapsed companion (Antokhin \etal 1982; Antokhin \& Cherepashchuk
1984), we discuss below  the implications of these observations with respect to this
scenario.
\subsubsection{The Expected Emergent X-ray Luminosity}
An interesting issue is whether the predicted X-ray luminosity produced
by the accretion process (after allowance for wind absorption) can be reconciled with the X-ray observations of
WR 134.

First, if we associate the 2.27 day periodicity with orbital motion,
we can obtain an estimate of the location of the compact object in
the WR wind. Assuming for simplicity a circular orbit (i.e., that the
circularization timescale is shorter than the evolutionary timescale; Tassoul 1990), a
canonical mass for the
secondary as a neutron star, $M_X$ = 1.4 M$_{\odot}$, and a mass for WR 134, $M_{\star}$ = 11
M$_{\odot}$ (Hamann, Koesterke, \& Wessolowski 1995), we obtain an
orbital separation of $a$ $\approx$ 17 R$_{\odot}$, or $a$ $\approx$ 6 WR core radii
(Hamann \etal 1995).\footnote{Since the
observationally determined masses of WN 6 stars in binary systems show a
large scatter ($M_{\star}$ $\approx$ 14 M$_{\odot}$ in WR 153; St-Louis \etal 1988
--- $M_{\star}$ $\approx$ 48 M$_{\odot}$ in WR 47; Moffat \etal 1990), we use
here a value given by atmospheric models of WR stars (adopting $M_{\star}$ $\approx$ 14 M$_{\odot}$ or 48 M$_{\odot}$ does not qualitatively change the conclusions presented in the following).}

Second, the {\it total} X-ray
luminosity (in ergs s$^{-1}$) produced by Bondi-Hoyle accretion of a stellar wind onto
a degenerate object can be expressed by the following relation (Stevens
\& Willis 1988):
\begin{equation}
L_X \approx 2.03 \times 10^{57} \: \eta \: \dot{M} \: M_X^2 \: a^{-2} \: v (a)^{-4} \left(1 - \frac{L_X}{L_E}\right)^2,
\end{equation}
where $\eta$ is the efficiency
of the conversion of gravitational energy into X-ray emission; $\dot{M}$ the
mass-loss rate of the primary (in M$_{\odot}$ yr$^{-1}$); $v (a)$ the WR wind velocity at the secondary's
location (in km s$^{-1}$), and $L_E$ the Eddington luminosity (in ergs
s$^{-1}$) given by
$L_E \approx 5.02 \times 10^{37} M_X/\sigma_e$ (the electron scattering
coefficient $\sigma_e$ = 0.35). In this
expression, we neglect the orbital
velocity of the secondary relative to the (much larger) WR wind
velocity. For the latter, we start by arbitrarily adopting the
well-established mean velocity law for O stars of the form (Pauldrach, Puls, \&
Kudritzki 1986)\footnote{We explore below the impact on  $L_X$ of more recently
proposed (but not yet well established) revisions for the velocity law
and mass-loss rate of WR stars.}:
\begin{equation}
v(r) = v_{\infty} \: (1 - R_{\star}/r)^{0.8}.
\end{equation}
Note that at the assumed location of the secondary ($r$ $\approx$ 6 $R_{\star}$; see above) the
wind has almost reached its terminal velocity,
$v_{\infty}$ = 1900 km s$^{-1}$ (Rochowicz \&
Niedzielski 1995). We further assume: $\dot{M}$ =
8 $\times$ 10$^{-5}$ M$_{\odot}$ yr$^{-1}$ (Hogg 1989), and $\eta$ = 0.1 (McCray
1977). This yields: $L_X$
$\approx$ 1.4 $\times$ 10$^{37}$ ergs s$^{-1}$.

We have modeled
the emergent X-ray flux by a power law with energy index $\alpha$, substituted above a characteristic high-energy cutoff $E_c$ by the function $\exp [(E_c -
E)/E_f]$, as commonly observed in accretion powered pulsars (White,
Swank, \& Holt 1983; Kretschmar \etal 1997). The values of $\alpha$,
$E_c$, and $E_f$ are chosen to be roughly representative of accreting
pulsars: $\alpha$ $\approx$ --
0.2; $E_c$ $\approx$ 15 keV; $E_f$ $\approx$ 15 keV (White \etal 1983). The X-ray spectrum was
normalized to the total X-ray luminosity determined above, i.e., $L_X$ =  1.4 $\times$ 10$^{37}$ ergs
s$^{-1}$.

The next step is to evaluate the attenuation of the beam of photons as they
propagate through the stellar wind. Since the absorption properties of
ionized plasmas are dramatically dependent on their ionization state (e.g.,
Woo \etal 1995), one has first to determine to what degree the
accreting compact object ionizes the surrounding stellar wind. The presence of an immersed, strongly ionizing companion
is expected to create an extended X-ray photoionized
zone in its vicinity (Hatchett \& McCray 1977). An investigation of the effect of the ionizing X-ray flux coming from the
secondary on the surrounding wind material requires a detailed
calculation of the radiative transfer (Kallman \& McCray 1982). However,
good insight into the ionization state of an \it optically thin \rm gas
illuminated by a X-ray point source can be obtained by considering the quantity
(Hatchett \& McCray 1977):
\begin{equation}
\xi(r, r_X) = \frac{L_X}{n(r)r_X^2} = \frac{4 \pi L_X \overline{m}}{\dot{M}} \: v(r) \left(\frac{r}{r_X}\right)^2,
\end{equation}
with: $n(r)$ the local number density of the gas; $r_X$ the distance
from the X-ray source. This expression is derived using the mass continuity
equation: $\dot{M}$ = 4 $\pi$ $r^2$ $\overline{m}$ $n(r)$ $v(r)$.
Here, $\overline{m}$ is the average mass per ion (we assume a pure helium atmosphere). The velocity of the material
as a function of $r$ is
given by equation (2). The variations of $\log \xi(r, r_X)$ are illustrated as
viewed from above the orbital plane in Figure 10. In the limit $\xi
\rightarrow 0$ ergs cm s$^{-1}$,
the ionization balance of the material is unaffected by the presence of
the X-ray emitter, i.e., is principally governed by the radiation field
of the WR star. We stress that this
model assumes an optically thin plasma. In reality, the mean path
length of the X-ray
photons is considerably shorter, and it is thus likely that the X-ray
photoionized zone would be dramatically less extended --- especially in
the direction toward the primary --- than sketched here. 

The predicted (after wind absorption) X-ray luminosities for the wavebands corresponding to the $PSPC$ and $IPC$ detectors onboard
the \it ROSAT \rm and \it Einstein \rm
observatories (0.2-2.4 and 0.2-4.0 keV, respectively) 
have been computed for three illustrative cases spanning a wide range in
the wind
ionization state, namely $\log \xi$ = 2.1, 1.8, and 0 ergs cm
s$^{-1}$. The photoelectric atomic
cross sections for ionized plasmas are taken from Woo \etal (1995). Thompson scattering by the free electrons was also taken into
account. The two former values, $\log \xi$ = 2.1 and 1.8  ergs cm
s$^{-1}$, are roughly what would be expected in the framework of our
model (Fig.10), whereas in the latter case, $\log \xi$ = 0 ergs cm
s$^{-1}$, we explore the (somewhat unrealistic) case in which the presence
of the neutron
star has no effect on the surrounding stellar wind. In this case, photoelectric cross
sections for cold interstellar gas were used (Morrison \& McCammon
1983). Because the WR wind material is, by nature, far from neutral, the
derived X-ray luminosities are here \it strictly lower \rm limits. 

Pollock (1987) reported on two pointed {\it IPC} observations of WR 134 obtained with the {\it Einstein} satellite, with no evidence for variability (see his Table 5). The derived (corrected for interstellar extinction) X-ray luminosity in the 0.2-4.0 keV band, 4.6 $\pm$ 1.6 $\times$ 10$^{32}$ ergs s$^{-1}$, being consistent with the upper limit derived by Sanders \etal (1985): 4.5 $\times$ 10$^{32}$ ergs s$^{-1}$ (when scaled to the same adopted distance of 2.1 kpc; van der Hucht \etal 1988). On the other hand, a single measurement (due to the {\it ROSAT} satellite) is available in the 0.2-2.4 keV range (Pollock, Haberl, \& Corcoran 1995). With no clear indication of X-ray variability, we will assume in the following that the {\it typical} observed (and corrected for interstellar extinction) X-ray luminosities from WR 134 are: 0.46 $\pm$
0.22 and 4.6 $\pm$ 1.6 $\times$ 10$^{32}$ ergs s$^{-1}$ in the 0.2-2.4 and 0.2-4.0 keV bands,
respectively.\footnote{Unfortunately, the
relatively large uncertainty in the period (\S 4.1.1) does not
allow to fold in phase the {\it ROSAT}
and {\it Einstein} data, thus preventing the construction of an observed X-ray light curve of WR 134.}

 These values can
be directly compared with the predicted X-ray luminosities after wind
absorption shown as a function
of orbital phase in Figure 11. For $\log \xi$ = 2.1 and 1.8 ergs cm
s$^{-1}$, the observed luminosities are systematically much lower than
expected in the framework of our model, with a deficiency reaching 2-3
orders of magnitude. For $\log \xi$ = 0 ergs cm
s$^{-1}$, the {\it ROSAT} data are consistent with the
expectations. However, and unless the {\it Einstein} satellite observed
WR 134 near X-ray eclipse, the same conclusion does not hold for the
data in the 0.2-4.0 keV band, with an order of magnitude
deficiency. Although mildly significant, this deficiency shows that, even in the extreme case in which the presence of
the accreting neutron star
has a negligible effect on the WR wind, the observed and predicted X-ray
luminosities can hardly be reconciled. Overall, and although we stress that
more detailed and rigorous calculations are necessary, the deficiency of the observed X-ray
flux must thus be regarded as serious. This conclusion is bolstered if
one considers that wind radiative instabilities may largely account for the observed fluxes (typically 10$^{32-33}$
ergs s$^{-1}$ in the 0.2-2.4 keV range; Wessolowski 1996).
 
Assuming a lower mass-loss rate due to
clumping (Moffat \& Robert 1994; Nugis, Crowther, \& Willis 1998) may, at best, reduce the discrepancy by
one order of magnitude (see equation [1]). In contrast, however, assuming a ``softer'' $v(r)$
law (i.e., $\beta$ $>$ 0.8; see equation [2]) for
WR stars (Schmutz 1997; L\'epine \& Moffat 1999; Antokhin,
Cherepashchuk, \& Yagola 1998), or taking into account
the existence of a velocity plateau inside the photoionized zone
(Blondin 1994), exacerbates the discrepancy.

Considering that the soft X-ray flux of WR
134 is very similar to that of other \it bona fide single \rm WN stars
(Pollock 1987), and was even  among the lowest found for single WN 6 stars
during the \it ROSAT \rm all-sky survey
(Pollock \etal 1995), this deficiency of predicted X-ray flux suggests centrifugal (or magnetic)
inhibition to prevent the gravitational
capture by the compact object of the primary's wind material (the
so-called ``propeller effect''; Lipunov 1982; Campana
1997; Cui 1997; Zhang, Yu, \& Zhang 1998). Note that orders of magnitude deficiency in the X-ray flux is observed 
in some high-mass X-ray binaries (HMXRBs; e.g., Taylor \etal 1996).
\subsubsection{Modeling the Line-Profile Variations Caused by an Ionizing X-ray Source}
Keeping in mind that the accretion process would be, as suggested above, much less
efficient than expected in the Bondi-Hoyle approximation --- and therefore
that the influence of the X-ray source on the surrounding wind material
would be far less severe than sketched in Figure 10 --- we performed
simulations of the LPVs caused by the orbital revolution of a localized, strongly
ionized wind cavity.

We proceed with numerical simulations of the 
observed LPVs of the representative He II \l4686 line, binning the epoch I
spectra to 0.1 phase resolution. We modify the SEI
code (Sobolev method with exact integration: Lamers, Cerruti-Sola, \& Perinotto
1987; hereafter LCP) to allow for the variation of the source function in a 3-D space, thus 
breaking the initially assumed spherical symmetry of the WR wind. The variation of the unperturbed source function as a function of the radial distance from the star is described by equation (4) of LCP. The wind dynamics is assumed to be
identical inside, and outside the photoionized cavity. Specifically, we adopt a $\beta$-velocity law with an exponent $\beta$ = 3 and a ratio of the wind velocity at the inner boundary of the optically thin part of the WR wind to the terminal velocity, $w_0$, of 0.25 (see equation [35] of LCP).

We start from the  assumption that the underlying, unperturbed line profile
is not affected by any phase-dependent variations. We fit this reference 
profile with the standard SEI code, preserving complete spherical symmetry. Throughout these simulations, we adopt $\tau_T$ = 0.60, $\alpha_1$ = 6.0, $\alpha_2$ = 0.5, $\epsilon_0^{\prime}$ = 12.0, $B_0$ = 1.2, $a_T$ = 0.2, and $w_D$ = 0.25 (these values are uncertain to about 30 \%). We refer the reader to LCP for a complete description of these parameters. Note that these values slightly differ from those adopted by Marchenko \& Moffat (1998b) due to a different approach in modelling (adopting their values, however, would not qualitatively modify the conclusions drawn in the following). Note 
that while the issue of the flattened, asymmetric wind in WR 134 is appealing
(Schulte-Ladbeck \etal 1992; Moffat
\& Marchenko 1993b), this inevitably introduces a great and probably
unnecessary complexity into the 
simulations. 

There are two obvious choices for the reference profile of the 
unperturbed wind: a minimum-emission or maximum-emission profile, derived from 
a smoothed minimum/maximum emissivity at a given wavelength for a given phase
(Fig.12). By introducing 3-D variations into the source function of the unperturbed 
wind corresponding to the minimum-emission reference profile, we immediately find 
that we are not able to reproduce the observed LPVs with any reasonable choice
of the parameters. Thus we adopt the maximum-emission representation of the 
unperturbed profile as a starting approximation. 

To simulate the X-ray induced 
cavity in the otherwise spherically symmetric wind, we require the following
free parameters: $\Theta$ --- the azimuthal extension of the cavity;
$\Delta z$ = $\Delta p$ --- line-of-sight and impact-parameter extension of the cavity; $\Delta z_0$ --- line-of-sight distance of the 
cavity's inner boundary from the stellar core; $i$ --- orbital inclination; $k$ ---
coefficient describing the deviation of the source function within the cavity, namely 
$S(cavity)$ = $k$ $\times$ $S(wind)$; $\phi_0$ --- cross-over phase (frontal passage 
of the cavity). We assume a circular orbit. A sketch of the adopted geometry is shown in Figure 13.

With these parameters, we attempt to fit all the phase-binned profiles, minimizing the 
observed minus model deviations in the $\chi^2$ sense, while reproducing 
the observed TVS of He II \l4686 (Fig.12). We also attempt to make
the (O -- C) profile deviations distributed as evenly as possible over the 
entire phase interval. We concentrate on the goodness of the fit for the 
velocities not exceeding $\pm$ 0.8 $v_\infty$, since it is impossible to match 
the red wing of the He II \l4686 profile (electron scattering effect;
see Hillier 1991) as well as the bluemost portion (partial blending with 
N III transitions). 

After an extensive search for optimal parameters, we find that the following set 
reproduces the observed LPVs reasonably well (Fig.14): 
$\Theta$ = 140 $\pm$ 10$^{\circ}$, $\Delta z$ = $\Delta p$ = 8 $\pm$
0.5 $R_{\star}$, 
$\Delta z_0$ = 2.7 $\pm$ 0.2 $R_\star$, $i$ = 65 $\pm$ 5$^{\circ}$, $k$ = 0.07 $\pm$ 0.01 (i.e., 
practically zero emissivity from the X-ray eroded wind), and $\phi_0$ =
0.55 $\pm$
0.03.  This value of $\phi_0$ means that the X-ray source passes in front
when the continuum flux undergoes a shallow minimum (Fig.6). The largest (O -- C) deviations are concentrated around phases 0.0 and
0.5. By no means are we able to reproduce the extended emissivity excesses
at $v$ $\approx$ -- 0.5 $v_\infty$ ($\phi$ = 0.35) and $v$ = + (0.5-0.7)
$v_\infty$ ($\phi$ = 0.65 and 0.75). These deviations \it might \rm be generated  by a low-amplitude 
velocity shift of the underlying profile as a whole, due to binary motion
(partially inducing the centroid variations seen in Fig.2). We will not consider this possibility further, as it will
add more free parameters into the model. We
are able to reproduce only the general appearance of the TVS as a structured, non-monotonic function with 3 distinct maxima, without precise match of their amplitudes and positions (Fig.12).

According to Model 5 of Kallman \& McCray (1982), helium is  fully
ionized for $\log \xi$ $\ga$ 1.5 ergs cm s$^{-1}$. The derived azimuthal
and radial extensions of the cavity indeed match fairly well the 
size of the highly ionized zone with $\log \xi$ $\approx$ 1.5 ergs cm s$^{-1}$
(Fig.10), especially if we take into account its large
azimuthal extension. However, since the accretion is likely to be partially
inhibited and the WR wind is not optically thin, the real photoionized
cavity will be much less extended than
sketched in Figure 10. For example, reduction of the efficiency of the X-ray
generation by 1-2 orders of magnitude would dramatically reduce the dimension of the $\log \xi$ $\approx$
1.5 cavity (to the size of the  $\log \xi$ $\ga$
2.5 cavities sketched in Figure 10; see equation [3]). The local nature of this putative zone is also suggested by an inspection of the P Cygni
absorption profile variations 
of He I \l4471 and N V \l4604. The epoch I
spectra have been groupped into broad phase bins of width 0.2 (the P Cygni absorptions are very weak,
thus requiring an extremely high S/N), and are overplotted in Figure 15. It is apparent that potential
phase-related variations do not significantly exceed the observationally-inflicted
accuracy.
Thus, neither the base of the wind (N V \l4604), nor the relatively distant 
regions (He I \l4471) are \it seriously \rm affected by the presence of the
photoionized cavity. An inspection of 16 archive \it IUE \rm SWP spectra of WR 134
obtained during the period 1989 November 30 - December 6 (see Table 3)
also shows
that no UV lines with well-developed P Cygni absorption components  show
any significant variability in their absorption troughs. Hints of relatively
weak, \it possibly \rm phase-locked variations are only evident in the
emission parts. 

We conclude that despite the encouraging similarity
between the observed and modeled line-profiles, the
relatively large
derived size of the cavity ($R$ $\approx$ 8 $R_{\star}$) along with its large azimuthal extension seem to enter in conflict with the low level of
X-ray flux observed from WR 134.
\subsubsection{On the Origin of the Long-Term Modulations}
Another question we may ask is whether this model is able to
account for the $\cal P$ $\approx$ 40 and $\cal P$ $\approx$ 625 day recurrence timescales present in the
photometric data (Marchenko \etal 1996b; Marchenko \& Moffat 1998a), as
well as the epoch-dependent nature of the LPVs (Figs.2 and 4).

As the wind approaches the compact object, it is
gravitationally focused, forming a standing bow
shock. This leads to the formation of an
extended wake downstream (Blondin \etal 1990). This process is suspected to be
non-stationary in HMXRBs, with the possible formation of a quasi-cyclical
``flip-flop'' instability (Benensohn, Lamb, \& Taam 1997; but see Ruffert 1997), with characteristic
timescales comparable to the flow time
across the accretion zone ($\approx$ minutes); too short a timescale to
account for \it any \rm observed long-term variations in WR 134.

Alternatively, it is tempting to associate the 40 day recurrence timescale
to the precession of a tilted accretion disk as suggested in some HMXRBs
(Heermskerk \& van Paradijs 1989; Wijnands, Kuulkers, \& Smale
1996). The condition for the creation of a persistent accretion disk around a wind-fed neutron star orbiting an early-type companion can be written as (Shapiro \& Lightman 1976):
\begin{equation}
2.7 \times 10^{-11} \gamma^8 B_{12}^{-4/7} R_{10}^{-10/7} M_X^{20/7} M_{\star}^{-4} P^{-2} R_{\star}^{4} L_{X}^{2/7} \ga 1
\end{equation}
With $B_{12}$ the surface magnetic field of the neutron star in units of 10$^{12}$ Gauss, $R_{10}$ the radius of the neutron star in units of 10 km, $M_X$ and $M_{\star}$ in units of solar masses, $P$ in units of days, $R_{\star}$ in units of solar radii, and $L_{X}$ in units of ergs s$^{-1}$. 
$\gamma$ is expressed by the relation: $\gamma \approx \zeta^{1/2} \left[v_{esc}/v(a)\right]$ where $\zeta$ is a dimensionless quantity accounting for the deviation from the Hoyle-Lyttleton treatment, and $v_{esc}$ the escape velocity at the WR ``photosphere'' (we neglect the orbital
velocity of the neutron star relative to the WR wind
velocity). Taking $B_{12}$ = $R_{10}$ = $\zeta$ = 1, we obtain a value much below unity ($\sim$ 10$^{-4}$), suggesting that  accretion mediated by a persistent disk in
WR 134 would be very unlikely.

A viable mechanism that might, however, account for the ~monthly changes
in the spectroscopic pattern of variability (Figs.2 and 4) is to consider long-term changes in the accretion rate (thus varying
the size of the ionized cavity surrounding the compact
companion).
\subsection{Wind-Related Variability?}
Although the
encouraging agreement between the observed and modeled LPVs (Fig.14)
leads to the conclusion
that the model involving an accreting compact companion cannot be completely
ruled out by the present observations, the very low observed X-ray flux of WR 134 which imposes inhibited
accretion, as well as the long-term
changes in the accretion rate required to account for the
epoch-dependent nature of the LPVs, set serious constraints on this model. This, together with the inconclusive search for
rapid \it periodic \rm photometric variations, possibly induced by a
spinning neutron star (Antokhin
\etal 1982; Marchenko \etal 1994) can be used to argue that the
existence of such a low-mass collapsed companion is
questionable.

Alternatively, it is conceivable that we observe 
rotationally-modulated variability in WR 134. Indeed, the global pattern of
spectral variability resembles in some aspects that of the apparently single WR
star EZ CMa, for which the existence of a rotation-modulated, structured wind has been proposed (St-Louis \etal 1995; Morel \etal
1997, 1998; Harries \etal 1999). By analogy with EZ CMa, the He II line profiles vary in a
fairly similar fashion (Vreux \etal 1992, and our Fig.9), changes in line skewness or FWHM generally
demonstrate a rather simple and well-defined behavior when phased with the
2.27 day period (Fig.2), and a positive/negative correlation between the pattern of
variability presented by different parts of the same line profile is occasionally
found (Fig.8). Other outstanding properties shared by WR 134 and EZ CMa are the
epoch-dependent nature of the variations, as well as the substantial
depolarization of the emission lines (Harries \etal 1998); this last point being generally attributed to an equatorial density enhancement (see also Ignace \etal 1998). 

Another
piece of evidence pointing to the similarity between WR 134 and EZ CMa
comes from the broadband ($\sim$ continuum) polarimetric observations of
Robert \etal (1989).  Plots of the Stokes parameters $Q$ and $U$ of WR 134 versus
phase are shown in Figure 16 (because of a likely loss of
coherency over long timescales, the two data subsets separated by a year
are plotted separately). The data show strong epoch-dependent
variations, with a clear single-wave variation in $U$ (less clear in $Q$) in
1985, and a fairly clear double-wave modulation in $Q$ (less clear in $U$)
in 1986.  This behavior is qualitatively very similar to what is
observed in EZ CMa (e.g., Robert \etal 1992), although the polarimetric
variations are not as clearly phase-locked in WR 134. Such a drastic change from a single-wave to a double-wave modulation in data
taken one year apart is not easily acommodated by a binary
hypothesis. However, neither the amount, nor the time coverage of the polarimetric data allow to regard it as a decisive evidence favoring the single-star interpretation. Although the 7
nights of linear spectropolarimetry of WR 134 by Schulte-Ladbeck \etal (1992) do show variations, both in the continuum and in the lines,
the data are sparsely spread out over 6 months, making it impossible to
separate variations on the 2.3-day cycle from long-term epoch-dependent
variations.  Certainly, further more intense spectropolarimetry on
different timescales in all 4 Stokes' parameters will prove quite
interesting. 

Relying on the
EZ CMa and WR 134 similarities, one may speculate that the 2.3 day
periodicity in WR 134 is
induced by  a small number of spatially
extended, relatively long-living and rotating wind streams whose formation is
possibly triggered by  photospheric perturbations, such as magnetic structures or pulsations
(e.g., Cranmer \& Owocki 1996; Kaper \etal 1997).
 Contrary to EZ CMa, however, WR 134 does not show any detectable phase-related 
changes in the UV or optical P Cygni absorption troughs, nor a clear
correlation between changes occuring near the hydrostatic stellar
``surface'' and in the wind.

By speculating about the existence of (non)radial pulsations in WR 134 (although a search for related rapid light variations
 has been hitherto inconclusive; Cherepashchuk 1975; Antokhin
\etal 1992), and exploring the idea of interactions between different pulsational modes, 
one is led to the notion of  quasi-periodic ejection of shell-like structures.
This idea can be explored via the application of the model
introduced above for the anisotropic wind ($\S$ 5.1.2). We allow for the formation of a 
shell with enhanced optical depth in the otherwise spherically symmetric wind. 
The shell geometry may be variable but, for the sake of simplicity, we mainly
explore the spherically symmetric case. The shell can slowly propagate outwards
at a given rate, slightly expanding and gradually approaching the physical 
conditions in the unperturbed surrounding wind. Indeed, at least the variations in the blue 
wing of He II \l4686 (Figs.4a and 14)
somewhat mimic the expected behavior of such an outwardly-moving
structure. However, there is also pronounced asymmetry between the blue
and red parts of the He II \l4686
profile, especially obvious around $\phi$ = 0.35-0.55 (Fig.14). This asymmetry 
cannot be reproduced by any adjustment of the model free parameters, unless 
one makes the contrived assumption that the front- and back-side lobes of the 
shell are formed under different physical conditions. This leads us to
discard the model of an expanding shell with central symmetry.

Although the existence of a globally inhomogeneous outflow in WR 134 is
attractive, such an interpretation is also not without difficulties. For instance, the origin of
the long-term (quasi)periodic changes in continuum flux, and of the epoch-dependency of the
spectral changes remains unexplained. Magnetic activity, by inducing
long-term changes in the global wind structure, may constitute a
convenient, although largely ad hoc, way to accomodate this aspect of
the variability.
\acknowledgments 
\noindent {\it Acknowledgments:} We wish to thank J. W. Woo for kindly
providing us with the photoelectric cross sections for ionized
plasmas. We acknowledge Alex W. Fullerton and the referee, Mike Corcoran, for their detailed and helpful comments. T. M., S. V. M., A. F. J. M., and N. S.-L. wish to thank the Natural Sciences
 and Engineering Research Council (NSERC) of Canada and the Fonds
pour la Formation de Chercheurs et l'Aide \`a la Recherche (FCAR) of Qu\'ebec for financial
 support. P. R. J. E. is grateful to CONACyT. I. I. A. acknowledges financial support from
Russian Foundation for Basic Research through the grants No. 96-02-19017 and
96-15-96489. T. E. is grateful for full financial aid from the
Evangelisches Studienwerk/Germany which is supported by the German Government.

\clearpage

\def\etal{{et al. \rm}}
\footnotesize \begin{center}
\rm TABLE 1 -- JOURNAL OF SPECTROSCOPIC OBSERVATIONS.
\end{center}  
\rm \begin{table}[h]
\centering
\begin{tabular}{llcccccc}
\hline \hline
 &      & \multicolumn{1}{c}{HJD}        &          & Number     & Spectral       & Reciprocal Dispersion 
     &     \\
\multicolumn{1}{c}{Epoch} & \multicolumn{1}{c}{Date} & -- 2440000 & \multicolumn{1}{c}{Observatory$^a$} & of Spectra & Coverage (\AA) & (\AA $\:$ pix$^{-1}$)  & S/N \\\hline
I  & 92 Jul - Aug   & 8813-8843   & SPM      & 65 & 4380-4790 & 0.41 & $\sim$ 160\\
II & 93 Oct   & 9258-9269   & SPM      & 38 & 3990-4840 & 0.83 & $\sim$ 115\\
III  & 95 May - Oct  & 9860-10005  & OMM, DAO & 32  & 4360-5080 & 1.62
& $\sim$ 210\\
    &      &             &          & 30 & 5050-5945 & 1.62 & $\sim$ 290\\\hline
\end{tabular}
\end{table}
$^a$ SPM: San Pedro
M\'artir Observatory 2.1 m (Mexico); OMM: Observatoire du Mont
M\'egantic 1.6 m (Canada); DAO: Dominion Astrophysical Observatory 1.2 m
(Canada).

\vspace*{1cm}

\begin{center}
\rm TABLE 2 -- JOURNAL OF PHOTOMETRIC OBSERVATIONS.
\end{center}  
\rm \begin{table}[h]
\centering
\begin{tabular}{lccccc}
\hline \hline
 & \multicolumn{1}{c}{HJD}           &             & Number of Binned      & Mode of       & Filter  \\
\multicolumn{1}{c}{Date} & -- 2440000        &  Observatory$^a$        & Observations      & Observation       & Set$^b$    \\\hline
89 Nov - 93 Feb & 7859-9045  & HIPPARCOS     & 44  & ---         & $Hp$ ($\sim$ $B$ + $V$)$^c$ \\
90 May - 97 Jul & 8025-10636 & APT           & 483 & Diff. phot. & $V$ \\
92 Jun - Jul & 8800-8828  & SPM           & 28  & 2-ch. phot. & 5185 (250) \\
92 Jun - Jul & 8800-8832  & SPM           & 23  & Diff. phot. & $V$ \\
92 Jul & 8819-8833  & MO            & 55  & Diff. phot. & 6012 (87) \\
95 Jun - Jul & 9885-9903  & CR            & 13  & Diff. phot. & $V$    \\
95 Jun - Aug & 9891-9949  & SPM           & 88  & Diff. phot. & $V$  \\\hline
\end{tabular}
\end{table}
$^a$ APT: Automatic Photometric Telescope 0.25 m (USA); SPM: San Pedro
M\'artir Observatory 0.84 m (Mexico); MO: Maidanak Observatory 0.6
m (Uzbekistan); CR: Observational Station in Crimea 0.6 m (Ukraine).\\
$^b$ For the midband filters, we provide the central wavelength and
FWHM (in \AA).\\
$^c$ See van Leeuwen \etal (1997). 

\clearpage

\begin{center}
\rm TABLE 3 -- IUE SWP HIGH RESOLUTION SPECTRA OF WR 134.
\end{center}  
\rm \begin{table}[h]
\centering
\begin{tabular}{ccc}
\hline \hline
SWP Image Number & JD -- 2440000$^a$ & Phase$^b$\\\hline
37705 & 7861.092 & 0.396\\
37707 & 7861.213 & 0.449\\
37718 & 7863.008 & 0.240\\
37725 & 7863.981 & 0.669\\
37727 & 7864.097 & 0.720\\
37734 & 7865.136 & 0.178\\
37735 & 7865.184 & 0.199\\
37737 & 7865.315 & 0.256\\
37739 & 7865.427 & 0.306\\
37741 & 7865.543 & 0.357\\
37743 & 7865.655 & 0.406\\
37745 & 7865.765 & 0.455\\
37748 & 7865.972 & 0.546\\
37750 & 7866.095 & 0.600\\
37755 & 7866.976 & 0.988\\
37757 & 7867.103 & 0.044\\\hline
\end{tabular}
\end{table}
\hspace*{4cm} $^a$ Julian Date at the midpoint of the exposure (2400 s for all
spectra).\\
\hspace*{4cm} $^b$ According to the ephemeris of McCandliss \etal (1994): $JD$
2447015.753 + 2.27 $E$.
\clearpage

\tighten
\def\etal{{et al. \rm}}
\def\l{$\lambda$}
\def\p{$\phi$ $\sim$ }
\def\km{km s$^{-1}$}

\begin{figure}[h]
\begin{minipage}{22cm}
\vskip -1.7truecm
\epsfxsize=16cm
\epsfysize=16cm
\epsfbox{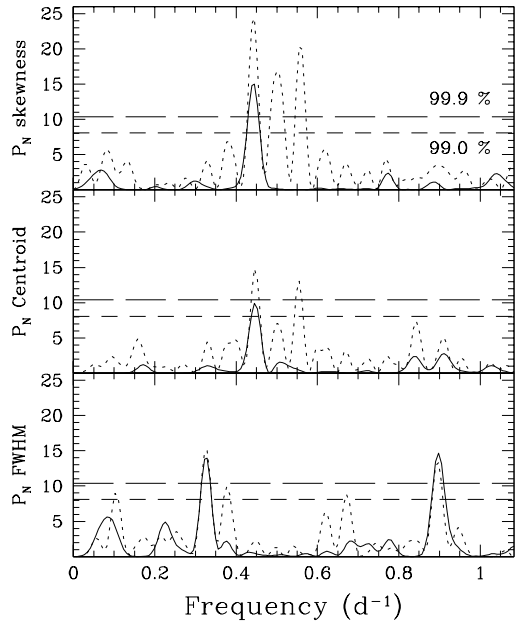}
\end{minipage}
\vskip 0.5truecm
\caption{Power spectra (normalized to the
total variance of the data) of the skewness, centroid, and FWHM time series of He
II \l4686 for epoch I. {\it Dotted line}: raw PS; {\it solid line}: CLEANed
PS. The short- and long-dashed horizontal lines indicate the
 99.0 \% and 99.9 \% confidence levels for the presence of a deterministic
signal in the time series, respectively. The number of iterations of the CLEAN algorithm, {\it N}, has been set to 3400 (skewness), 380 (centroid), and 4500 (FWHM); a gain {\it g} = 0.2 has been used throughout. The algorithm is generally insensitive to the combination of these two parameters in the range {\it N} $\ga$ 100 and  {\it g} $\leq$ 0.5-0.8.}
\end{figure}
 
\clearpage

\begin{figure}[h]
\begin{minipage}{22cm}
\vskip -1.7truecm
\epsfxsize=16cm
\epsfysize=16cm
\epsfbox{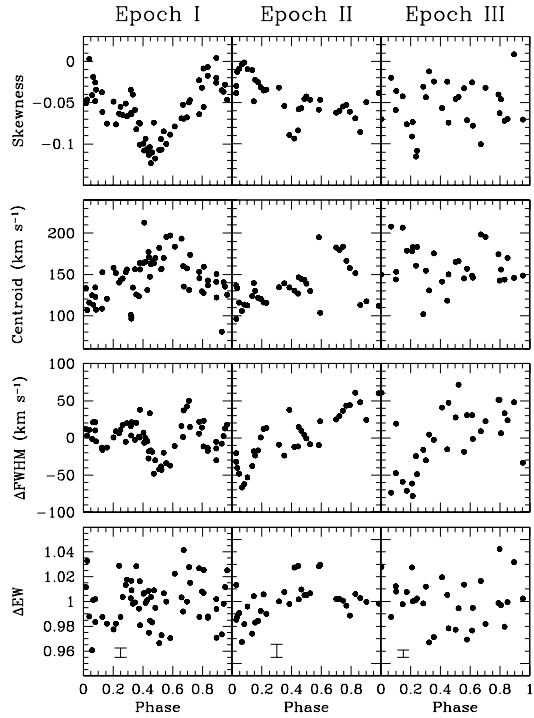}
\end{minipage}
\vskip 0.5truecm
\caption{Skewness and centroid variations, deviations
of the FWHM around the mean value (in \km), and EW variations
(normalized by division to
the mean value) of He
II \l4686 for each epoch, as a function of phase. The 2-$\sigma$ error
bars for the EW values were calculated
according to Chalabaev \& Maillard (1983). As everywhere in this
paper, the ephemeris
of McCandliss \etal (1994) has been adopted: $HJD$ 2,447,015.753 + 2.27 $E$.}
\end{figure}
 
\clearpage

\begin{figure}[h]
\begin{minipage}{22cm}
\vskip -1.7truecm
\epsfxsize=16cm
\epsfysize=16cm
\epsfbox{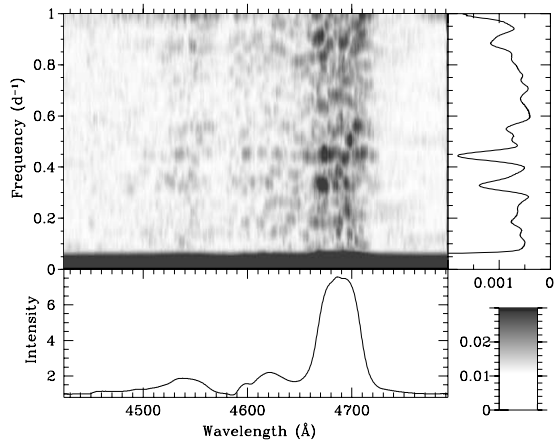} 
\end{minipage}
\vskip 0.5truecm
\caption{The pixel-to-pixel CLEANed PS of the rectified spectra of epoch I. The lower and right-hand panels show the
mean spectrum and the normalized, integrated power over all pixels, respectively. The highest peak corresponds to
a frequency: $\nu_0$ $\approx$ 0.444 d$^{-1}$. The gain and the number of iterations have been set to 0.3 and 250, respectively.}
\end{figure}
 
\clearpage

\begin{figure}[h]
\epsfbox{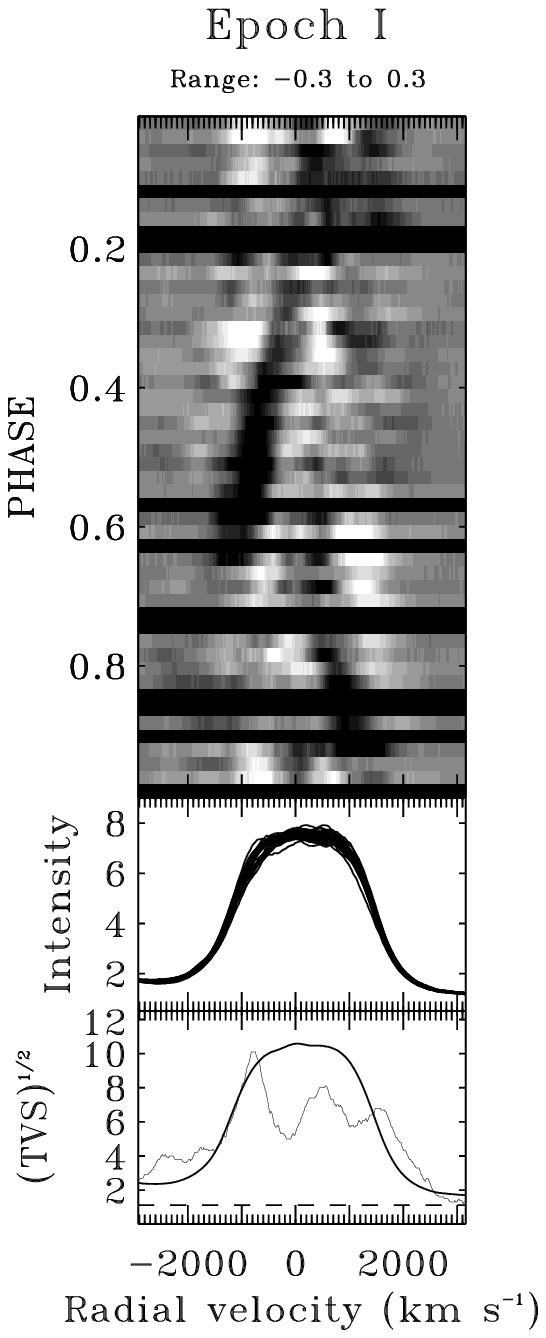}
\vskip -13.2truecm
\epsfbox{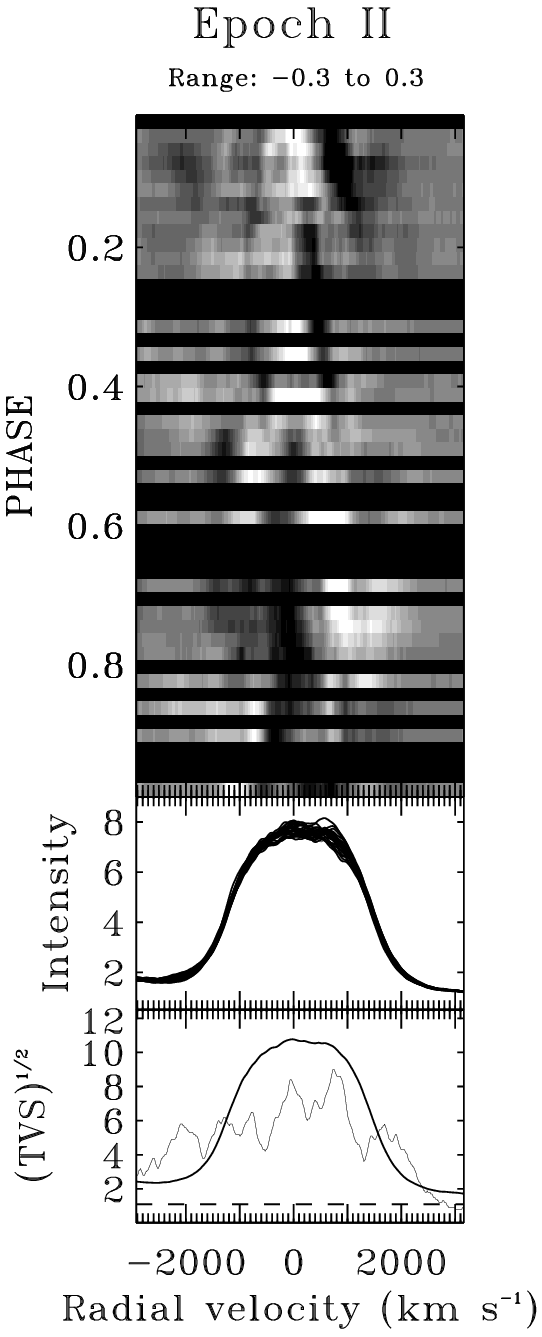}
\vskip -25.5truecm
\epsfbox{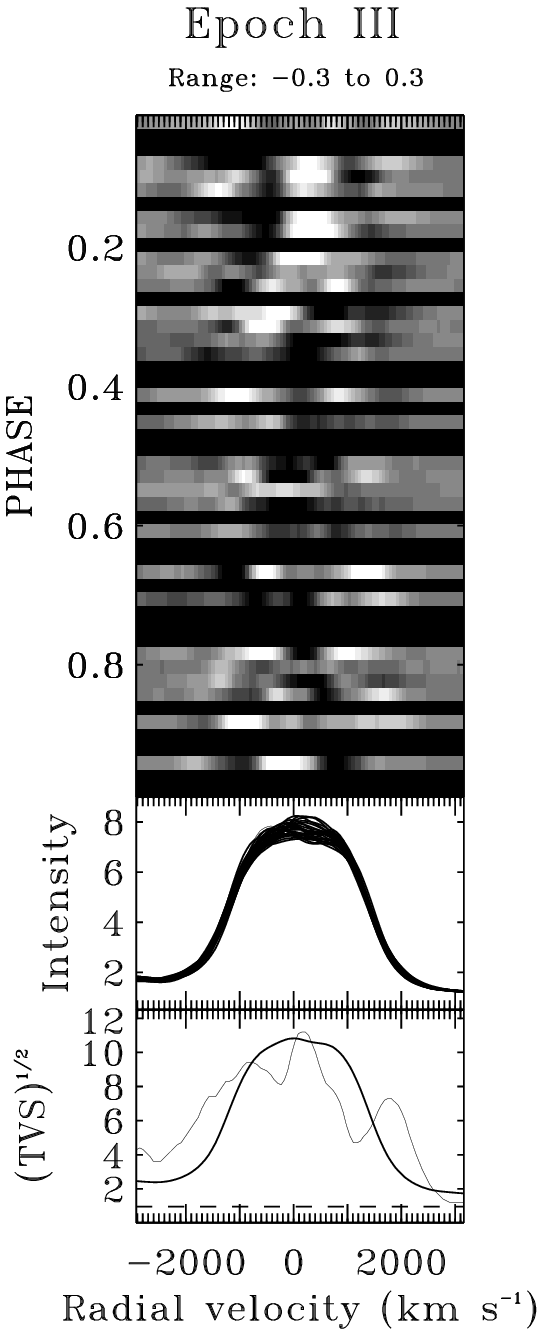}
\vskip -17.5truecm
\caption{Gray-scale plots of the time series
of the residuals of He II \l4686 for each epoch, as a function of
phase. The
residuals are the binned (to 0.02 phase
resolution) He II \l4686 line profiles  minus their
corresponding unweighted means for each epoch. Excess emission components appear
brighter in these plots. The middle portion of each panel shows a
superposition of the rectified profiles. The values of the temporal variance spectrum (TVS;
Fullerton, Gies, \& Bolton 1996), along with the horizontal dashed
line indicating the 99.0 \%
confidence level for significant variability, are displayed in the
lower portion of each panel. The mean profile (in arbitrary units) of the epoch is overplotted. The
gray-scales plots are displayed in the radial velocity frame
(the radial velocities are referred to the line laboratory rest wavelength).}
\end{figure}
 
\clearpage

\begin{figure}[h]
\begin{minipage}{22cm}
\vskip -1.7truecm
\epsfxsize=16cm
\epsfysize=16cm
\epsfbox{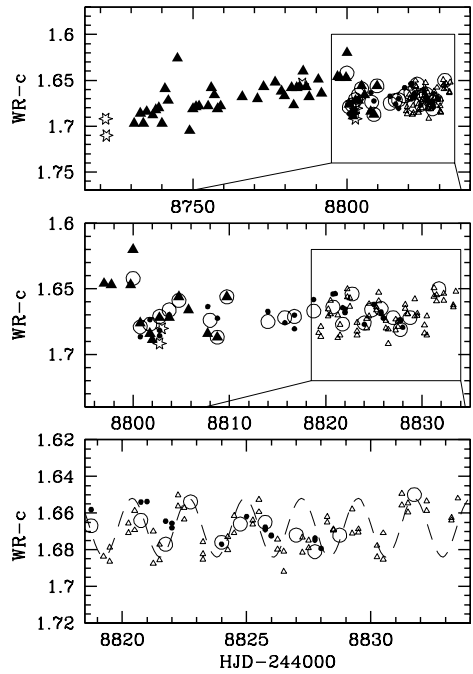}
\end{minipage}
\vskip 0.5truecm
\caption{Photometry of WR 134 in 1992. {\it Filled
triangles}: APT; {\it open circles}: SPM (Mexico) 1-channel
photometry; {\it filled circles}: SPM  2-channel photometry; {\it open
triangles}: Maidanak Observatory; {\it open stars}: $HIPPARCOS$. The bottom
panel is an enlargement of the insert in the central panel. The sizes of the symbols correspond to 2-$\sigma$
error bars. In order to guide the eye in the bottom frame, a sinusoid
with a period of 2.27 days (ephemeris of McCandliss \etal 1994) has
been overplotted.}
\end{figure}
 
\clearpage

\begin{figure}[h]
\begin{minipage}{22cm}
\vskip -1.7truecm
\epsfxsize=16cm
\epsfysize=16cm
\epsfbox{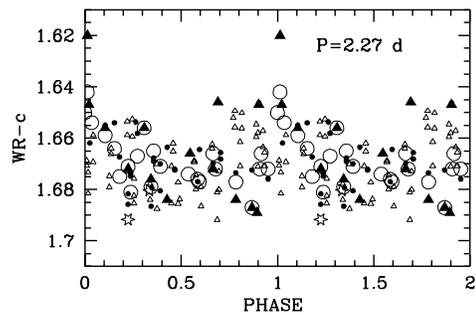}
\end{minipage}
\vskip 0.5truecm
\caption{Photometry of WR 134 in 1992, folded with the
$\cal P$ = 2.27 day period. Only the data plotted
in the middle panel of Figure 5 have been used.
The symbols are coded  as in Figure 5.}
\end{figure}
 
\clearpage

\begin{figure}[h]
\begin{minipage}{22cm}
\vskip -1.7truecm
\epsfxsize=16cm
\epsfysize=16cm
\epsfbox{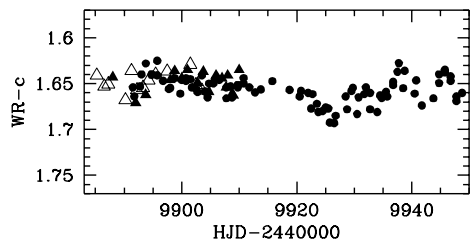}
\end{minipage}
\vskip 0.5truecm
\caption{Photometry of WR 134 in 1995. {\it Filled
triangles}: APT;  {\it filled circles}: SPM (Mexico); {\it open
triangles}: Crimean Observational Station. The sizes of the
symbols correspond to 2-$\sigma$ error bars.}
\end{figure}
 
\clearpage

\begin{figure}[h]
\begin{minipage}{22cm}
\vskip -1.7truecm
\epsfxsize=16cm
\epsfysize=16cm
\epsfbox{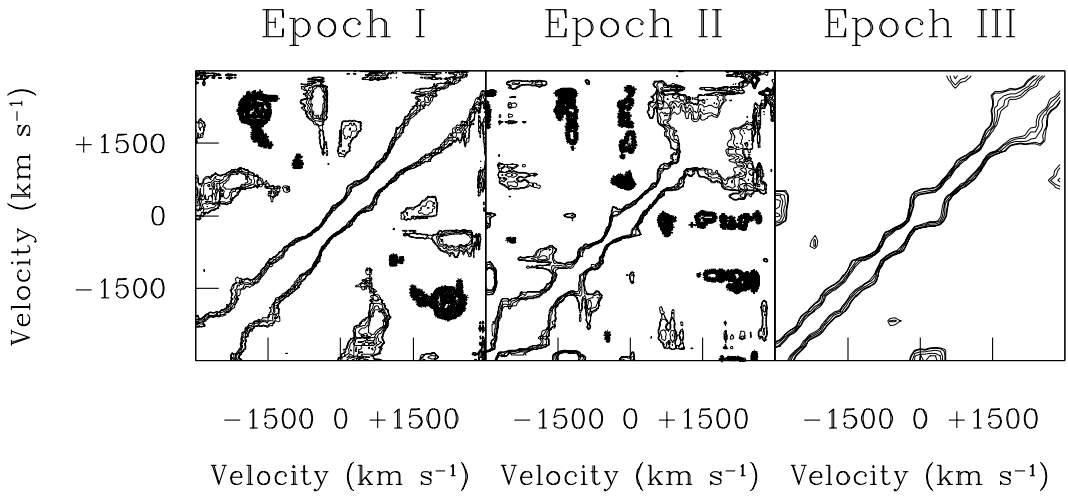}
\end{minipage}
\vskip 0.5truecm
\caption{Auto-correlation matrices of He II
\l4686 for each epoch. Thick and thin contours indicate a negative or positive correlation in the pattern of variability presented by the
 same line profile at different  radial velocities, respectively. The
lowest contour is drawn for a significant correlation at the 99.9 \% confidence level.}
\end{figure}
 
\clearpage

\begin{figure}[h]
\begin{minipage}{22cm}
\vskip -1.7truecm
\epsfxsize=16cm
\epsfysize=16cm
\epsfbox{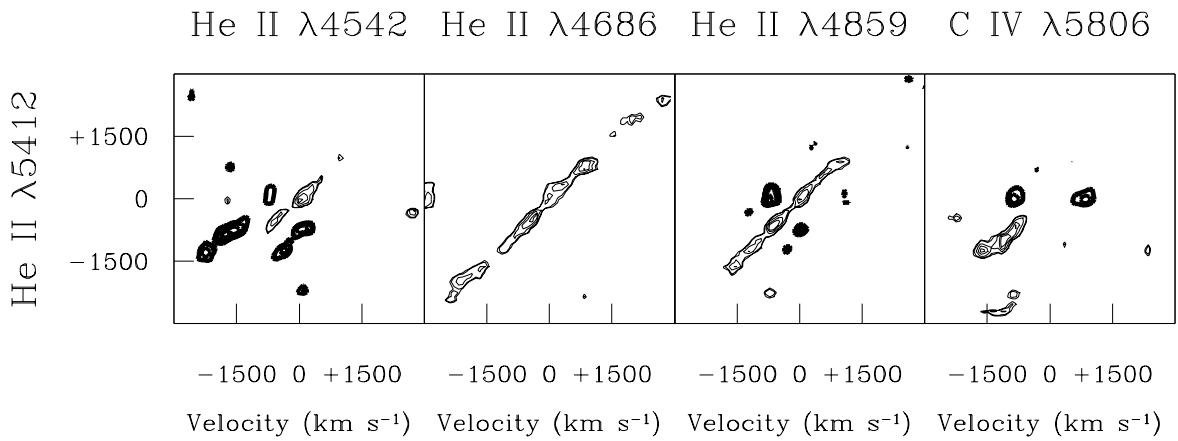}
\end{minipage}
\vskip 0.5truecm
\caption{Correlation matrices of He II
\l5412 with He II \l4542, He II \l4686, He II \l4859, and C IV \l5806
for epoch III.  Thick and thin contours indicate a negative
or positive correlation in the pattern of variability presented by two
 line profiles at different  radial velocities (referred to the line laboratory rest wavelength), respectively. The
lowest contour is drawn for a significant correlation at the 99.9 \% confidence level.}
\end{figure}
 
\clearpage

\begin{figure}[h]
\begin{minipage}{22cm}
\vskip -1.7truecm
\epsfxsize=16cm
\epsfysize=16cm
\epsfbox{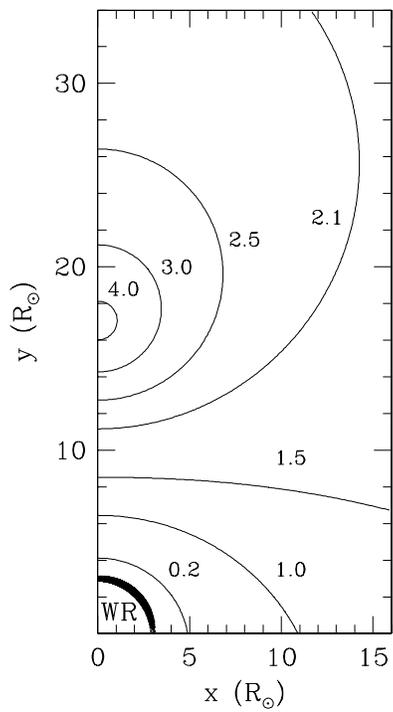}
\end{minipage}
\vskip 0.5truecm
\caption{Contours of constant ionization parameter
$\log \xi$ (in ergs cm s$^{-1}$). The accreting, strongly ionizing compact
companion is located at (0, 17). The WR star is placed at the origin of this plot.}
\end{figure}
 
\clearpage

\begin{figure}[h]
\begin{minipage}{22cm}
\vskip -1.7truecm
\epsfxsize=16cm
\epsfysize=16cm
\epsfbox{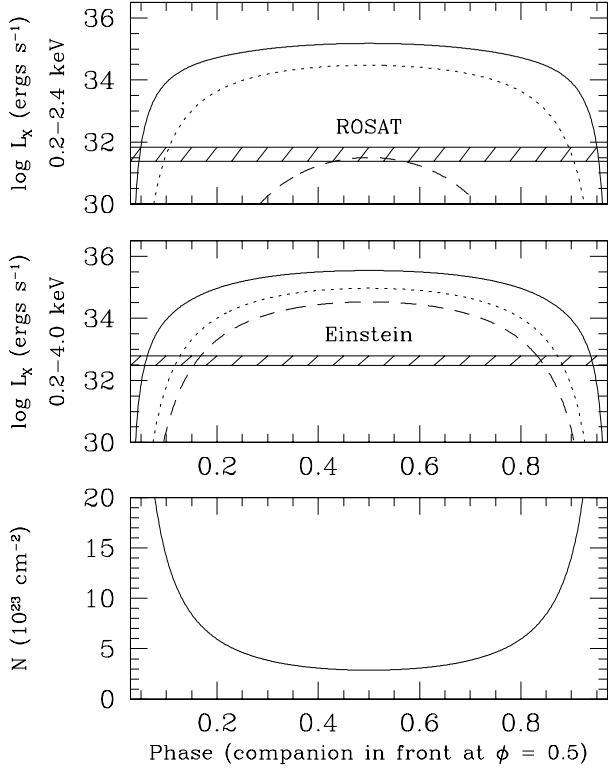}
\end{minipage}
\vskip 0.5truecm
\caption{{\it Upper and middle panels}: predicted X-ray luminosities after allowance for wind absorption in the 0.2-2.4
and 0.2-4.0 keV bands as a function of the orbital phase $\phi$, and for different values of
the ionization parameter $\xi$: $\log \xi$ = 2.1 ergs cm s$^{-1}$ ({\it solid line}),
$\log \xi$ = 1.8 ergs cm s$^{-1}$ ({\it short-dashed line}), and $\log \xi$ = 0 ergs cm
s$^{-1}$ ({\it long-dashed line}). The shaded aeras show the range of
the {\it ROSAT} and {\it Einstein} count rates for WR 134, namely 0.46 $\pm$
0.22 and 4.6 $\pm$ 1.6 $\times$ 10$^{32}$ ergs s$^{-1}$ in the 0.2-2.4 and 0.2-4.0 keV bands,
respectively (Pollock, Haberl, \& Corcoran 1995; Pollock 1987). The luminosities have been scaled to a distance of 2.1 kpc (van der Hucht \etal 1988). Note that the
calculations do not consider the X-ray emission intrinsic to the WR
wind (generally ascribed to radiative instabilities), nor the scattering of the neutron
star emission in the WR wind which may cause extra emission at
X-ray eclipse. {\it Lower panel}:
column density of the absorbing wind material in units of 10$^{23}$ cm$^{-2}$, as a
function of $\phi$.}
\end{figure}
 
\clearpage

\begin{figure}[h]
\begin{minipage}{22cm}
\vskip -1.7truecm
\epsfxsize=16cm
\epsfysize=16cm
\epsfbox{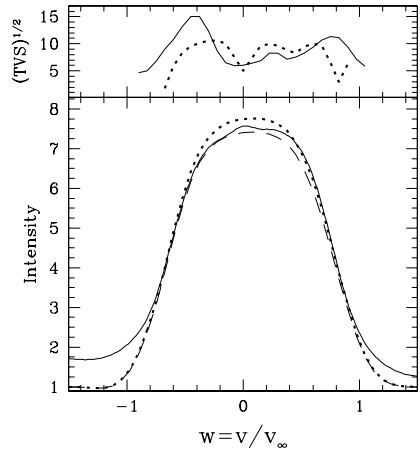}
\end{minipage}
\vskip 0.5truecm
\caption{{\it Upper panel: solid line} --- the observed TVS; {\it dotted line} --- the modeled
TVS within the maximum-emission approach. {\it Lower panel: solid line} --- mean epoch I profile of 
He II \l4686; {\it dotted line} --- the SEI model fit within the maximum-emission 
approach; {\it dashed line} --- the SEI model fit applying the minimum-emission 
approach. Note that the model is unable to fit the bluemost and redmost
parts of the line profile because of partial blending and electron
scattering effect, respectively. The projected velocity, {\it w}, is normalized to $v_{\infty}$ = 1900 \km.}
\end{figure}
 
\clearpage

\begin{figure}[h]
\begin{minipage}{22cm}
\vskip -1.7truecm
\epsfxsize=16cm
\epsfysize=16cm
\epsfbox{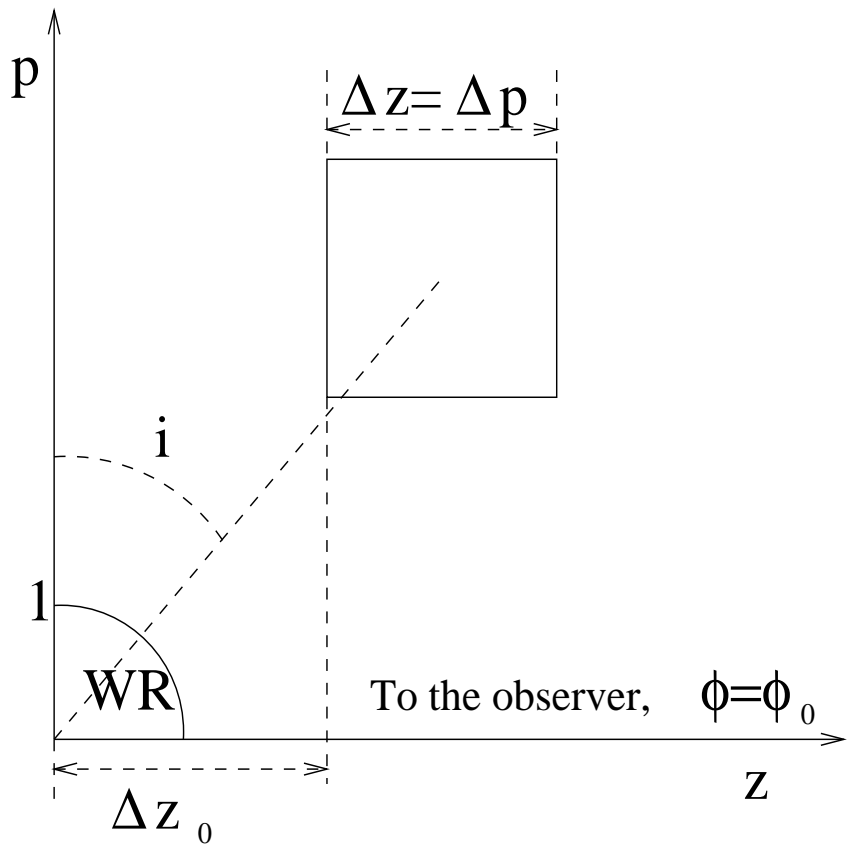}
\end{minipage}
\vskip 0.5truecm
\caption{Sketch of the geometry adopted for the photoionization cavity. 
The azimuthal extension of the cavity, $\Theta$, is in the direction perpendicular to the 
plane of the figure.}
\end{figure}
 
\clearpage

\begin{figure}[h]
\begin{minipage}{22cm}
\vskip -1.7truecm
\epsfxsize=16cm
\epsfysize=16cm
\epsfbox{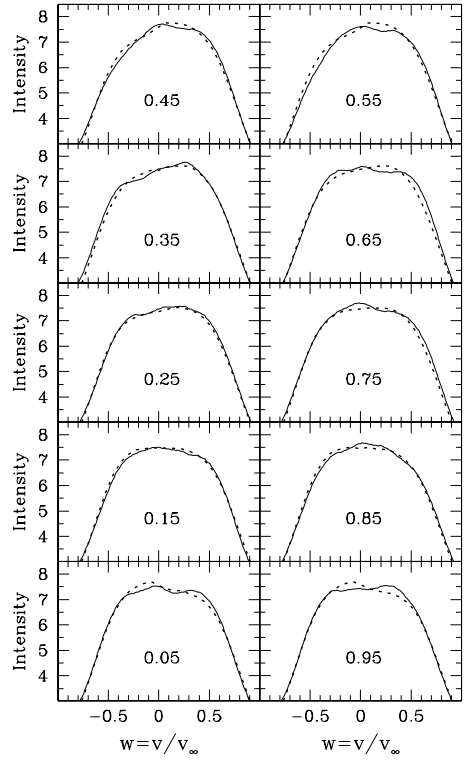}
\end{minipage}
\vskip 0.5truecm
\caption{The observed ({\it solid line}) and modeled ({\it dotted line})
phase-dependent variations of He II \l4686. All epoch I spectra are 
binned to 0.1 phase resolution; the bin mid-phase is shown in each
panel.}
\end{figure} 
 
\clearpage

\begin{figure}[h]
\begin{minipage}{22cm}
\vskip -1.7truecm
\epsfxsize=16cm
\epsfysize=16cm
\epsfbox{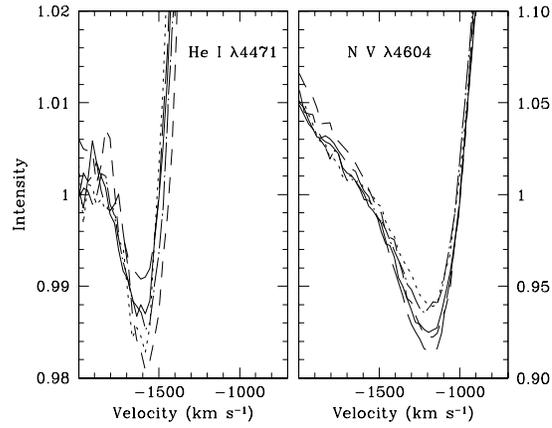}
\end{minipage}
\vskip 0.5truecm
\caption{Variations of the P Cygni absorption troughs 
of He I \l4471 and  N V \l4604 with phase. All epoch I observations 
are binned to 0.2 phase resolution: {\it dashed-dotted line}: $\phi$ =
0.15-0.35; {\it long-dashed line}: $\phi$ = 0.35-0.55; {\it short-dashed line}:
$\phi$ = 0.55-0.75; {\it dotted line}: $\phi$ = 0.75-0.95; {\it solid line}: $\phi$
= 0.95-1.15. Note that the intensity scale is different for the two panels.}
\end{figure}  

\clearpage

\begin{figure}[h]
\begin{minipage}{22cm}
\vskip -1.7truecm
\epsfxsize=16cm
\epsfysize=16cm
\epsfbox{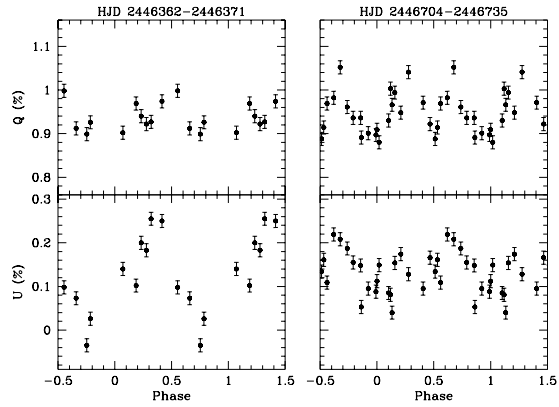}
\end{minipage}
\vskip 0.5truecm
\caption{The Stokes parameters $Q$ and $U$ of WR 134
in 1985-1986 (Robert \etal 1989) plotted in phase according to the
ephemeris of McCandliss \etal (1994). {\it Left and right panels}: data (with 2-$\sigma$ error bars) for
the intervals HJD 2,446,362-2,446,371 (1985) and HJD 2,446,704-2,446,735 (1986),
respectively. }
\end{figure}  
\end{document}